\documentclass{elsart1p}
\newcommand{\met}{\hbox{E\kern-0.5em\lower-0.1ex\hbox{/}}_T}
\newcommand\simlt{\lower.5ex\hbox{$\; \buildrel < \over \sim \;$}}
\newcommand\simgt{\lower.5ex\hbox{$\; \buildrel > \over \sim \;$}}

\usepackage{graphicx}
\usepackage{amssymb}
\begin{document}
\begin{frontmatter}
%
%
%
\title{Probing Cosmic Accelerators Using VHE Gamma Rays and UHE cosmic rays}
%
%
\author{Amir Levinson}
\address{Raymond and Beverly Sackler School of Physics \& Astronomy, Tel Aviv University, Tel Aviv, Israel}
\begin{abstract}
The $\gamma$-ray emission observed in several classes of Galactic and
extragalactic astrophysical sources appears to be linked to accreting
black holes and rotational powered neutron stars.   These systems are prodigious cosmic accelerators, and
are also potential sources of the UHE cosmic rays detected by several
experiments and VHE neutrinos.  We review a recent progress in
our understanding of these objects, and demonstrate how recent and future
observations can be employed to probe the conditions in the sources.
\end{abstract}
\end{frontmatter}
%
\section{Introduction}
\label{}
The recent launch of Fermi/GLAST and the rapid development of ground based experiments to measure TeV rays 
(HESS, VERITAS, MAGIC, HAWC), 
VHE neutrinos (ICECUBE, km3net) and UHE cosmic rays (Pierre Auger observatory) mark the 
beginning of a new era in high-energy astrophysics.   
To date, around 80 objects have been reported as VHE ($>100$ GeV) $\gamma$-ray sources, among which are 
active galactic nuclei (AGNs), pulsar wind
nebulae (PWNs), supernova remnants (SNRs), Galactic $\gamma$-ray binary systems, and a stellar cluster. 
Several thousands sources are expected to be detected by Fermi, the majority of which are AGNs, but also
a large number of gamma-ray bursts (GRBs) and pulsars.

Besides providing valuable information on the nature of VHE astrophysical systems, the various experiments described
above can advance our understanding of fundamental physics, e.g., strong general relativistic (GR) effects, supercritical 
magnetic fields, neutrino oscillations; probe new physics, e.g., violation of Lorentz invariance;
and can be employed to extract information about the Universe.  A brief account of some recent developments  
in VHE astronomy is given below. 
 
\section{Gamma ray sources}
The majority of $\gamma$-ray sources discovered thus far appear to be associated with compact 
astrophysical systems.
The common view is that the $\gamma$-ray emission observed is produced in fast outflows
expelled by a compact, central engine.  The compact source may involve a magnetized neutron star,
as in the case of pulsars, $\gamma$-ray binaries, and magnetars, or an accreting black hole system involving
either a supermassive black hole, as in blazars, or a stellar mass black hole as in microquasars and some classes of GRBs.  
VHE emission is expected also from other systems, e.g., clusters of galaxies, but will not be discussed below.

\subsection{Accreting black hole systems}
In accreting black hole systems the outflows are ejected along the polar region and appear to be
relativistic and highly collimated. Acceleration and collimation seem to occur on rather small scales, of the order of $10^2 r_g$
in M87, where $r_g=2GM/c^2$ is the Schwarzschild radius of the putative black hole. The collimation may be accomplished 
through magnetic hoop stresses or the pressure and inertia 
of a surrounding matter, and may provide an important dissipation channel of the bulk energy, even 
on relatively small scales \cite{brom07,Sik08}.  AGNs, microquasars and long duration GRBs are examples of $\gamma$-ray 
emitters that are thought to be powered by accreting black holes.  Despite the apparent differences in some basic 
parameters, most notably the black hole mass, accretion rate, and Lorentz factor of the outflow, 
the different classes of sources mentioned above share the same physics.  There are some important 
environmental differences that need to be carefully accounted for when confronting models with the data.
In the collapsar scenario for long GRBs \cite{Mcw99} the outflow emerges from a dense stellar envelope that extends 
to large radii $\sim 10^5r_g$ and can affect the outflow and its emission.  In microquasars involving a 
high mass companion, such as the TeV microquasar Cygnus X-1, the relativistic jet interacts with the companion's wind 
and radiation (e.g., Ref~\cite{brk09} 
and references therein).  This can potentially lead to modulations of the VHE emission and other complexities. 
Clear orbital modulations of the TeV flux have been detected in the TeV binary LS 5039 \cite{ahr06}, however, 
the nature of the compact object in this system is controversial \cite{brk09,rom07}, as discussed further below. 

The best constraints on the properties of the central engine and on jet kinematics have been obtained for 
some TeV AGNs.  Opacity arguments applied to several TeV blazars that exhibit rapid TeV variability
implies large Doppler factors, in excess of 30 in most of these sources \cite{lev06,beg08}, in conflict with the much lower
values measured for motions of radio patterns on VLBI scales.  Some explanations for this apparent discrepancy have been 
offered (e.g., Ref.~\cite{lev07} and references therein).  Recent observations 
suggest that the blazar zone may be located at radii much larger than previously thought, in which case
reconsideration of some aspects is needed \cite{brom09,Sik08}.  Detection of rapid variability of the resolved X-ray emission
from HST-1 in M87 indicates that large amplitude variations can somehow occur at a distance from the putative 
black hole which is three orders of magnitude larger than the inferred size of the emission region \cite{cheu07}. 
This has interesting implications for the hydrodynamics of the outflow that has been considered recently \cite{brom09,staw06}.

Even more interesting is the extreme TeV flares reported for PKS 2155-304
and Mrk 501 \cite{ahr07,alb07}. Since the characteristic size of any disturbance produced by the central engine cannot 
be smaller than size of the engine, $\sim r_g$, as measured in the rest frame of the engine, the associated variability 
time, if produced by dissipation of a significant fraction of the ejected energy, cannot be much shorter 
than $r_g/c$, regardless of Doppler boosting. The few minuets doubling time reported for PKS 2155-304 
then naively implies a black hole mass $M_{BH}<5\times10^7M_\odot$ \cite{lev08b,der08} in this source, in marked contrast to 
the value $M_{BH}=2\times10^9M_\odot$ inferred from the black hole-bulge relation \cite{ahr07}.  
The variability may imprint the scale of some external disturbance, e.g., in the case of a collision of
a fluid shell expelled from the central source with some intervening cloud of size $d<<r_g$.  However, 
this would imply correspondingly larger power for the event since only a fraction $(d/r_g)^2$ of the bulk 
power released by the engine can be tapped to produce the observed flare.  Note that the power extracted magnetically 
from the black hole scales as $L_{BZ}\propto r_g^2 B^2$, so that it automatically accommodates the requirement on the 
engine power \cite{lev08b}.  The question remains as to what is the fate of the remaining energy.  It could be 
that it dissipates at much larger radii over much longer timescales.  Any of the above scenarios implies
radiative inefficient accretion near the Eddington rate during this extreme episode \cite{lev08b}

It has been proposed that the variable TeV emission seen in M87 and perhaps some other blazars may originate from
the BH magnetosphere in a ``pulsar like'' process \cite{lev00,nero07}.  If true then it may provide a probe of 
the structure of the inner black hole magnetosphere.  To avoid vacuum breakdown and excess $\gamma\gamma$ opacity 
requires very low disk luminosity.  This requirement is consistent with observations of M87 and perhaps some TeV  
blazars.  The inferred jet power in those objects implies that the accretion energy is released predominately 
in mechanical or electromagnetic (i.e., Poynting flux) form rather than radiation.  
One might worry about other opacity sources that can absorb the TeV photons emitted from the magnetosphere in TeV blazars, in 
particular the synchrotron photons produced in the jet.  The observed 
synchrotron flux in this scenario should be produced at sufficiently large radii to allow escape of the TeV photons.
This ``pulsar like'' mechanism is also relevant for UHECRs production and is 
discussed further below in some detail.

\subsection{Pulsars and pulsar wind nebulae}
Pulsars exhibit both a polar outflow and a striped equatorial wind.  Gamma ray emission 
can in principle be produced in an inner gap located in the polar cap, in an outer gap, and in regions
where the equatorial wind dissipates \cite{hard07,hard05}.  Emission from the pulsar 
magnetosphere (the inner and/or outer gap)
should be modulated with a period that corresponds to the pulsar rotation period.  Several $\gamma$-ray pulsars
have been detected by EGRET, most of which have counterparts in other bands, but with a phase difference. 
Recently it has been reported \cite{abodo08} that Fermi detected a $\gamma$-ray pulsar which appears to have no counterpart,
and it could well be that many of the EGRET unidentified sources are in fact pulsars.  The characteristics of 
the $\gamma$-ray emission (e.g., upper cutoff of the spectrum) can be used to probe the magnetosphere and distinguish
between different models.  Emission from the pulsar wind nebula can reach very high energies, in excess 
of 100 TeV in the case of the Crab nebula.  It can be used to study the properties of the wind and its dissipation.

In binary systems involving a massive companion the emission 
originates from regions where the pulsar wind is shocked by the wind emanating from the companion star, as
in PSR B1259-63. The VHE emission from those regions is modulated mainly by the changing opacity along the sight line,
owing to the orbital rotation \cite{brk09}. 
The lack of accretion features in the X-ray spectrum of LS 5039 and LS I +61 303, and the interesting 
morphology of the extended radio emission in the latter object, is the basis for the claim
that these two TeV microquasars are in fact $\gamma$-ray binaries rather than accreting black hole 
systems.  This issue is currently under debate.  In particular, one naively expects similar effects 
from the interaction of a jet with a dense stellar wind. 
 
\subsection{Supernova remnants}
Supernovae blast waves have long been considered to be the prime sources of cosmic rays (CRs) at energies below the knee 
($\sim 10^{15}$ eV), and in some scenarios even up to $10^{18}$ eV, depending mainly on the strength
of the magnetic field in the acceleration zone.  An important diagnostic of CR production
in SNRs is VHE emission via inelastic collisions of the accelerated CRs with the ambient matter.  Much efforts 
have been devoted to search for this emission with the various TeV observatories.
Thus far 8 SNRs have been detected at TeV energies.  Whether the observed emission 
is of hadronic or leptonic origin is under debate \cite{ahr06b,port06,katz08}.  In particular, radio and X-ray 
observations seem to limit the contribution of hadronic emission to the TeV flux measured in SNRs RX J1713.7-3946
and RX J6852.0-4622 to a few percents at most \cite{katz08}.

\section{Using VHE $\gamma$-ray sources to probe the Universe}
Absorption of VHE $\gamma$-rays by the extragalactic background light (EBL) causes strong attenuation
of the flux emitted from distant sources.  This, on the one hand, complicates the interpretation of 
observational data.  On the other hand, it can be used to constrain the EBL.  In fact, the spectra 
of TeV blazars, that in some cases extends beyond 10 TeV, have already been used to place stringent
constraints on the EBL and, as a consequence, on the cosmological evolution of galaxies \cite{prim08} .  

Absorption of VHE $\gamma$-rays by the infrared background light can also be used, in principle, to 
impose constraints on the extragalactic magnetic field.  The idea is to detect extended, secondary emission
around blazars, resulting from cooling of e$^+$e$^-$ pairs that are produced via initiation of pair 
cascades in the intergalactic space by absorption of the primary TeV photons emitted from the source.
The properties of such $\gamma$-ray halos should reflect the strength of the extragalactic magnetic field.

\section{Astrophysical sources of UHECRs}
The origin of cosmic rays is yet an open issue.  The spectrum of cosmic rays extends over 
many decades in energy.  As stated above, the spectrum below the knee is most likely 
produced in Galactic supernovae.  Cosmic rays in the range $10^{15}-10^{18}$ eV may also 
be accelerated in regular SNRs\footnote{The change in the slope across the knee may then be due to propagation 
effects.}, provided that sufficiently strong magnetic fields ($\sim$ mG) can be somehow generated
in the shock region.  
Alternatively, this component may be produced in trans-relativistic blast waves \cite{bud08} or
may have extragalactic origin, though the smooth transition at the knee seems somewhat unnatural in the latter case. 

The origin of UHECRs (those above the ankle) remains a mystery.  It is widely believed that the
sources are extragalactic, though they have not yet been identified.
There is some evidence for a weak anisotropy in the
arrival directions of UHECRs events \cite{Auger08} that suggests a correlation of
the UHECRs sources with the large-scale structure in the local Universe \cite{Kash08}.
A general constraint on UHECRs sources can be derived from the requirement that the accelerated particles
are confined to the acceleration region; specifically that the escape time $t_{\rm esc}=r/c\Gamma$ is longer
than the acceleration time $t_{\rm acc}\simeq r_{L}(\epsilon)/c$, where $r_L(\epsilon)$ is the Larmor radius of
a particle having energy $\epsilon$.  This gives a relation between the source size and the strength of magnetic
field that depends to some extent on the composition of UHECRs.  Under the assumption that the 
UHECRs are accelerated in a relativistic magnetized outflow this also implies a minimum outflow power: 
\begin{equation}
L_j>10^{46} \Gamma^2\left(\frac{\epsilon}{10^{20} {\rm eV}}\right)^2 \qquad {\rm erg\ s^{-1}} .   
\label{Lj}
\end{equation} 
If the outflow is magnetically driven, then its power scales roughly as 
$L_{BZ}=10^{47}\epsilon B_4^2M_9^2$  erg s$^{-1}$, where  $M_9=M_{BH}/10^9M_{\odot}$ is the black hole mass in 
fiducial units and  $B=10^4B_4$ G is the strength of the magnetic field threading the horizon.  The 
efficiency factor $\epsilon$ depends on the geometry of the magnetic 
field and other details, and typically $\epsilon\simlt 0.1$.  So in principle supermassive black holes with
$M_9\simgt1$, $B_4\simgt1$ and stellar sources with $M_9\sim 10^{-9}$, $B_4\simgt10^{11}$ can account
for the required power.

The properties of the UHECRs sources are further constrained by X-ray and $\gamma$-ray surveys.  Based on a comparison
of the observed number density of active flares in various high-energy bands with the number density of 
sources required to account for the UHECRs production rate, it has been argued \cite{wax09} that if a fraction $\eta>10^{-2}$ 
of the total source power $L_j$ is converted to electromagnetic radiation, then the sources 
must be transients with $L_j>10^{50}$ erg s$^{-1}$.  The most natural candidates in this 
case are GRBs.  Alternatively, the sources must have very low radiative efficiency.  This is particularly 
relevant for models in which UHECRs are accelerated in AGN jets.  An example of such a ``dark'' source is M87, 
that exhibits a jet with an estimated power of $L_{j}>10^{44}$ erg s$^{-1}$, which is larger 
than the bolometric luminosity by a factor of at least $10^3$.  However, the jet power
is not large enough to satisfy the confinement limit. 

The condition (\ref{Lj}) should not necessarily apply in cases where the UHECRs are accelerated in regions 
that violate ideal MHD.  A specific 
example, originally due to Boldt and Ghosh \cite{BG99}, is starved black hole magnetospheres in dormant AGNs.
The maximal electric potential difference that can be generated by a rapidly rotating black hole is
\begin{equation}
\Delta V\sim 4.4\times10^{20} B_4M_9 \qquad {\rm volts}. \label{DeltV}
\end{equation} 
For $M_9\simgt1$, $B_4\sim1$ this potential is marginally sufficient to account for the 
highest UHECRs observed.   This full potential drop can be retained provided vacuum breakdown 
is prevented.
Breakdown of the gap will occur under conditions that allow formation of intense pair cascades.
This, however, requires a rather strong magnetic field and does not seem to be a major concern \cite{lev00}.
Alternatively, vacuum breakdown may occur through the agency of ambient radiation. To avoid this requires   
the luminosity of ambient radiation to satisfy $L<10^{38}M_9 \tilde{R}^2(\epsilon_s/0.1 {\rm eV})$ erg         
s$^{-1}$, where $\tilde{R}$ is the size of the external radiation source in units of $r_g$ and $\epsilon_s$
its peak energy \cite{lev00}.

As shown in Ref.~\cite{lev00} the accelerated particles
suffer severe energy losses, owing to curvature radiation, that limit their energy to values somewhat smaller than 
the maximum potential drop given in Eq. (\ref{DeltV}).   The peak energy of the curvature photons
depends on the charge $Z$ of the accelerated particle but not on its mass, and is typically of the order of a few TeV.
The total power released in the form of UHE particles and curvature 
photons depends on the rate at which charged nuclei are injected into the magnetosphere.  To account for the observed 
flux of UHECRs only a small fraction of the total Blandford-Znajeck power available is required on the average from 
a single source, consistent with the assumption that the parallel electric field in the gap is 
unscreened \cite{lev08b}.  Nonetheless, 
for sources within the GZK sphere the predicted flux of curvature TeV photons is above detection limit of current 
experiments, at least for some.  

Associating the UHECRs with either, ``dark'' AGN jets or starved black holes in dormant AGNs has interesting
implications for accretion.  The point is that the large magnetic field strength required in the vicinity of the 
horizon in either scenario, $B>10^3$ G, implies near Eddington accretion.  Since no bright quasars exist 
within the GZK sphere, this in turn implies radiative inefficient accretion, as seems to be the case in M87.


%
%

%
\end{document}